\documentstyle[editedvolume,psfig]{crckapb}   

\begin{opening}
\title{A Morphological Classification Scheme for ULIRGs}

\subtitle{Evidence for Multiple Mergers}

\author{K.D. Borne}
\institute{Raytheon Information Technology and Science Services\\
           NASA Goddard Space Flight Center, Greenbelt, MD USA}
\author{H. Bushouse}
\author{L. Colina}
\author{R.A. Lucas}
\institute{STScI, Baltimore, MD USA}
\author{A. Baker}
\author{D. Clements}
\institute{Cardiff University, Wales UK}
\author{A. Lawrence}
\institute{Institute for Astronomy\\
           University of Edinburgh, Scotland UK}
\author{S. Oliver}
\author{M. Rowan-Robinson}
\institute{Blackett Laboratory \\
           Imperial College, London, England UK}

\end{opening}

\runningtitle{Morphological Classification of ULIRGs}

\begin{document}


\section{Introduction}

The Hubble Space Telescope (HST) has been used to study a large sample
of ultraluminous IR galaxies (ULIRGs).  
With a rich legacy database of
$\sim$150 high-resolution images, we are studying
the fine-scale structure of this unique collection of violently
starbursting systems (\citeauthor{Borne97a}~1997a,b,c,d).
We review here some of the latest results from our survey.

\section{An HST Imaging Survey}

Our combined data set includes $\sim$120 WFPC2 I-band (F814W)
images and $\sim$30 NICMOS H-band (F160W) images.  These are
being used for multi-color analyses over a significant wavelength
baseline.  The NICMOS images are used specifically to probe through
some of the dust obscuration that plagues the shorter wavelength
images.  The full set of images is being used to study the galaxies'
cores and starburst regions.  Nearly all ULIRGs show
evidence for a recent tidal interaction, and we are identifying
the merger progenitors near the center of each galaxy, a task made
significantly easier with the H-band images.  These images are proving
to be particularly useful in mapping the spatial distribution of
starburst activity in these galaxies, in verifying the presence or
absence of a bright active nucleus, in deriving the distribution (in
both size and luminosity) of the multiple nuclei seen in these galaxies,
and in determining if these multiple cores are
the merger's remnant nuclei or super star clusters
formed in the merger/starburst event.

\section{HST Results and Serendipity}

\begin{figure}[ht]
\centerline{\psfig{file=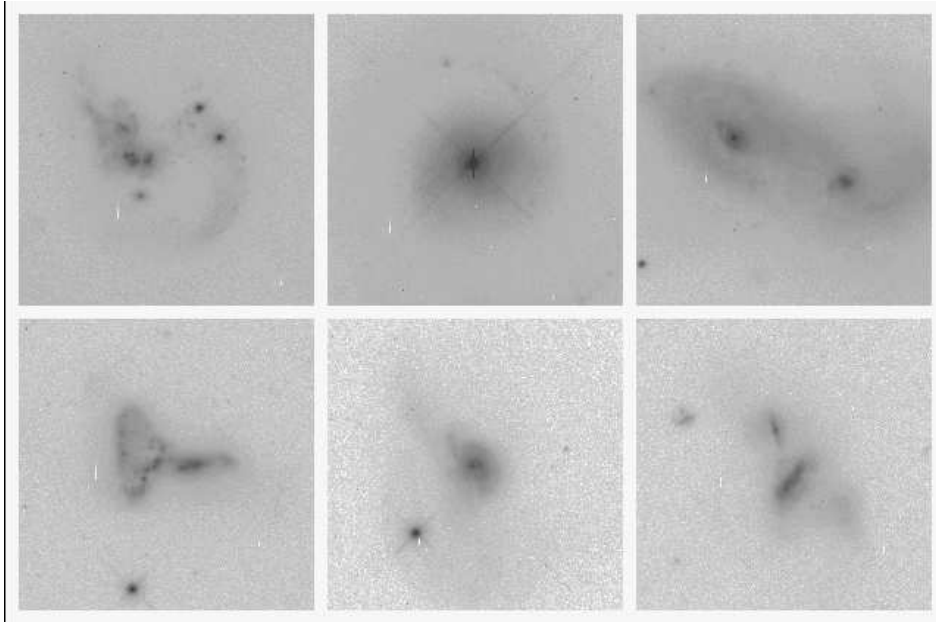,width=\textwidth,angle=-90}}
\caption{
HST I-band images of 6 representative ULIRGs.
{\bf (a)~Upper Left} -- IR1858$+$6527 --
A significant fraction of the ULIRGs appear
very disturbed, though primarily showing evidence for only a single
remnant galaxy (presumably the remnant of a major merger event).
In many of these cases, massive
star formation is seen on all scales in the HST images, including super 
star clusters of the type seen in HST images of other
colliding galaxies, indicating that a giant
starburst is the dominant power source in most ULIRGs.
{\bf (b)~Upper Middle} -- IR00509$+$1225 --
A star--like nucleus is seen in
15\% of the ULIRGs, for which the dominant power source may be a
dust--enshrouded AGN/QSO.
{\bf (c)~Upper Right} -- IR19254$-$7245 --
Many ULIRGs show evidence for strong
interactions among two or more galaxies.
Several of the galaxies show clear evidence in HST images for
a ring around the central nucleus (as shown here).
These are very similar to the rings seen in HST images
around the centers of other `black hole'--powered galaxies (e.g., NGC 4261).
{\bf (d)~Bottom Left} -- IR21130$-$4446 --
There is evidence for {\it{at least}} one classical
collisional ring galaxy in the sample, similar to the Cartwheel ring
galaxy imaged by the HST.
{\bf (e)~Bottom Middle} -- IR1717$+$5444 --
Some ULIRGs previously classified as
non-interacting from ground-based images now show in HST images
clear evidence of merging (a second nucleus) or 
of interaction (e.g., tidal tails).
{\bf (f)~Bottom Right} -- IR1353$+$2920 --
Many ULIRGs appear to have 
physically associated companion galaxies, which
may be related to the
collision, merger, and subsequent burst of star formation.
In these cases, the signs of interaction (e.g., distortions) are 
sometimes weak.
} \label{kdb:figure1}
\end{figure}

Several new discoveries have been made through our HST imaging survey
of the ULIRG sample (see Fig.~\ref{kdb:figure1} for some
representative images).  A few specific results are presented
in the following discussion.

\subsection{Mrk 273 = IR13428$+$5608}

A strong dust lane and a system of extended filaments have been 
discovered near the center of Mrk 273.  
The filaments are similar to those
seen in M82 and are probably indicative of a strong outflow 
induced by a massive starburst.  The central
region of the galaxy contains several separate cores, which may be
remnant cores from more than 2 merging galaxies.

\subsection{The SuperAntennae = IR19254$-$7245 }

One of the most interesting ULIRGs is IR19254$-$7245
(the SuperAntennae; \citeauthor{Mirabel91}~1991).   
With a morphology similar to
the Antennae galaxies (N4038/39; \citeauthor{Whitmore95}~1995), 
it is clearly the result of a collision between
two spirals.  In the case of the SuperAntennae,
the tidal arms have a total end-to-end extent of 350 kpc   
--- 10 times larger than the Antennae!  We have
resolved the two galaxies' nuclei (8$''$ separation) 
and have discovered a small
torus with diameter $\approx 2''$
around the center of each galaxy (Fig.~\ref{kdb:figure1}$c$).  
The southern component
is known to have an active nucleus and the torus may be related to the
AGN.  It is possible
that the northern galaxy also hosts an AGN, but the active nucleus
is obscured from view by a large column of dust.
Follow-up higher-resolution images with the HST PC 
have revealed a double-nucleus at the center of 
{\it{each galaxy}}, clearly suggesting a 
{\it{multiple-merger origin}}
for the SuperAntennae.

\subsection{Interaction / Merger Fraction}

Given the high angular resolution
($\sim$0.1--0.2$''$) of our HST images, a number of ULIRGs that
were previously classified as ``non-interacting'' have now revealed
secondary nuclei at their centers (remnant nuclei from a merger event?)
and additional tidal features (tails, loops).  
An example of one such system is shown in Figure~\ref{kdb:figure1}$e$.
It now appears that
the fraction of ULIRGs that show evidence for interaction is very
close to 100\%.  Observational estimates of 
this number have varied from 30\% to 100\% over the
past 10 years, but it now seems to be converging on a value significantly
above 90\%
(as indicated in the early work by \citeauthor{Sanders88}~1988).

\subsection{AGN Fraction}

The most significant question about the ULIRG phenomenon is
the nature of the power source.  That power source is generating
the ultra-high IR luminosities ($L_{\tt{IR}} > 10^{12}L_\odot$) through
dust heating and the corresponding
conversion of UV/optical radiation into IR radiation.
\citeauthor{Veilleux97}~(1997) have shown that the frequency of AGN-powered
ULIRGs increases sharply at $L_{\tt{IR}} \geq 10^{12.3}L_\odot$.
It is very likely then that a combination of starburst power and AGN
power is responsible for the ULIRG phenomenon among the various
galaxies comprising the whole sample,
and it is even possible that both power sources contribute energy
in unique proportions within each individual ULIRG.  In the latter
scenario, the power source for
the higher-luminosity ULIRGs is mainly the
AGN and for the lower-luminosity ULIRGs (still quite luminous)
it is the starburst.  
We have noted a particular morphological tendency in our HST images:
objects whose nuclei appear most star-like (i.e., unresolved)
also seem to be those that have been classified (from ground-based
{\it{spectroscopic}} observations) to be AGN.
About 15\% of our total sample have unresolved nuclei
(similar to the AGN fraction found by \citeauthor{Genzel98}~1998, 
and others).
This may represent the true fraction of ULIRGs that are dominated
by an AGN power source.  In the other cases, the observed
near-IR flux (in HST images) is clearly
spatially distributed among numerous bright star-forming (starbursting)
knots, which therefore are very likely the primary energy sources
for dust-heating.

\section{Morphological Classification of ULIRGs}

We have examined a complete subsample of ULIRG images
and have identified 4 main morphological classes, plus 2 additional
sub-classes (which are included in the main classes for 
statistical counting purposes).
Figure~1 depicts 6 representative ULIRGs, 
one from each of these classes:

\begin{enumerate}

\item Strongly Disturbed Single Galaxy (Fig.~1$a$)

\item Dominant AGN/QSO Nucleus (Fig.~1$b$)

\item Strongly Interacting Multiple-Galaxy System (Fig.~1$c$)

\item Weakly Interacting Compact Groupings of Galaxies (Fig.~1$f$)

\item Collisional Ring Galaxy (Fig.~1$d$)

\item Previously Classified Non-Interacting Galaxy (Fig.~1$e$)

\end{enumerate}

\begin{table}[htb]
\begin{center}
\caption{ULIRG Morphological Classes}
\begin{tabular}{lcccl}  
\hline
Class & Number & Fraction & $<log L_{IR}/L_\odot>$ 
& Notes \\
\hline
Disturbed Singles     & 30      & 34\%         & 11.85 & morph. class 1 \\
AGN/QSO Nucleus       & 13      & 15\%         & 11.74 & morph. class 2 \\
Interacting Multiples & 29      & 33\%         & 11.81 & morph. class 3 \\
Compact Groupings     & 14      & 16\%         & 11.94 & morph. class 4 \\
Collisional Rings     & 1-3     & $\sim$1-3\%  & ...   & re-classified \\
``Non-Interacting''   & $\sim$5 & $\sim$5\%    & ...   & re-classified \\
\hline
\end{tabular} 
\end{center}
\end{table}

\noindent
We show in Table~1 the
distribution of ULIRGs among these morphological classes.
It is seen here that there is little luminosity dependence among
the classes and that there is a roughly equal likelihood
that a ULIRG will appear either single (classes 1 and 2) 
or multiple (classes 3 and 4).

\section{Evidence for Multiple Mergers}

\begin{figure}
\centerline{\psfig{file=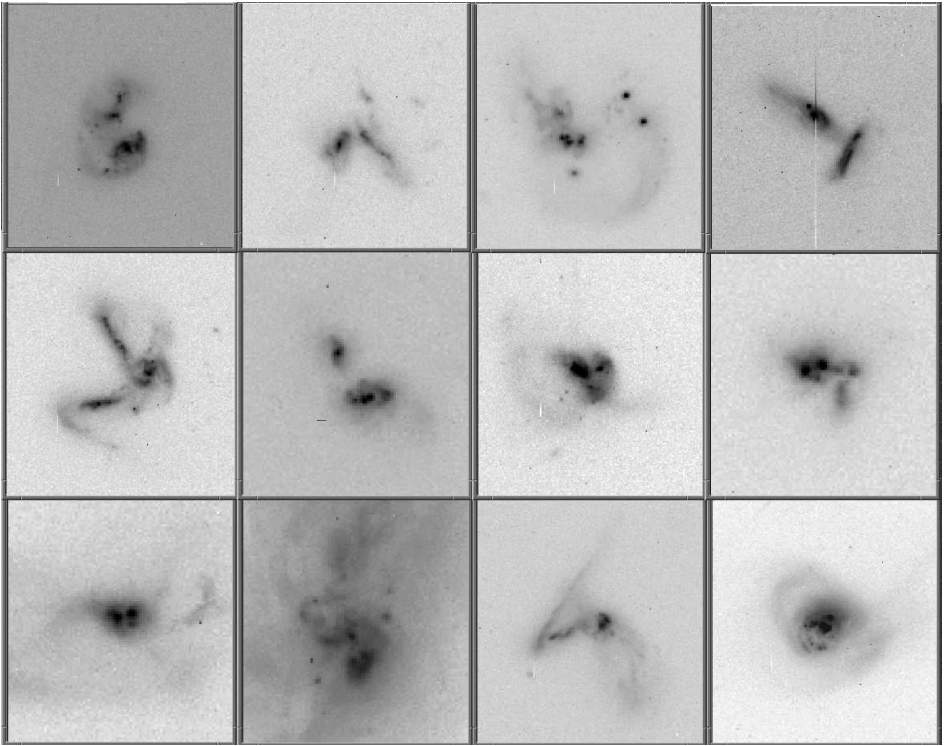,width=\textwidth,angle=-90}}
\caption{
Sample of ULIRGs whose morphologies appear to be derived from $>1$ merger.
} \label{kdb:figure2}
\end{figure}

Many of the recent results on ULIRGs are pointing to a complicated
dynamical history.  It is not obvious that there is a
well-defined dynamical point during a merger at which the ULIRG phase
develops, nor is it clear what the duration of the ultraluminous phase
is.  Our new HST imaging surveys indicate that the mergers are well
developed (with full coalescence) for some ULIRGs.  Others show clear
evidence for 2 (or more) nuclei.  While others ($\sim$5\%)
can best be described as wide binaries, still a long way from
coalescence.  One possible explanation for this 
{\it{dynamical diversity}} has been proposed
recently by \citeauthor{Taniguchi98}~(1998).  They
suggested a multiple-merger model for ULIRGs.  In this scenario, the
existence of double nuclei is taken as evidence of a second merger,
following the creation of the current starburst nuclei from a prior set
of mergers.  In fact, this would indicate for some systems (with double
AGN or double starburst nuclei) that the currently observed merger is
the third (at least) in the evolutionary sequence for that galaxy.  This
may seem unrealistic, but it may not be so unreasonable
if these particular ULIRGs
are {\it{remnants of previous compact groups of galaxies}}.  Compact groups
are known to be strongly unstable to merging, and yet examples are seen
in the local (aged) universe.  These may be the tail of a distribution
of dynamically evolving galaxy groups.  Similarly, the ULIRGs are
presumed to be historically at the tail end of a distribution of major
gas-rich mergers.  A connection between the two populations, if only in
a few cases, is therefore not unreasonable.  Figure~\ref{kdb:figure2}
presents images of 12 ULIRGs from our HST sample that appear to have
evolved from multiple mergers.  The evidence for this includes:
$>$2 remnant nuclei, or $>$2 galaxies, or an overly complex
system of tidal tails, filaments, and loops.

\section{Acknowledgments}

Support for this work was provided by NASA through
grant numbers GO--6346.01--95A and 
GO--7896.01--96A from the Space Telescope Science Institute, which
is operated by AURA, Inc., under NASA contract NAS5--26555.

{}

\end{document}